\documentclass[proceedings]{JHEP}

\usepackage{amsmath,amssymb,cite,epsfig,graphicx,epic,eepic}

\conference{International Europhysics Conference on High Energy
Physics (EPS HEP 2001)}

\title{T-violating effects in three flavor neutrino oscillations in matter}

\author{Tommy Ohlsson\thanks{Talk presented at the International
Europhysics Conference on High Energy Physics (EPS HEP 2001),
Budapest, Hungary, July 12-18, 2001. In collaboration with: Evgeny
Akhmedov, Patrick Huber, and Manfred Lindner. To be published in the
Conference Proceedings section of the Journal of High Energy Physics
(JHEP).}\\ Institut f{\"u}r Theoretische Physik, 
Physik-Department, Technische Universit{\"a}t M{\"u}nchen\\
James-Franck-Stra\ss{}e, DE-85748 Garching bei M{\"u}nchen, Germany\\
E-mail: {\tt tohlsson@ph.tum.de}}

\abstract{In this talk, we consider the interplay of fundamental and
matter-induced T-violating effects in neutrino
oscillations in matter. We present a simple approximative analytical
formula for the T-violating probability asymmetry for three flavor
neutrino oscillations in matter with an arbitrary density profile. We
also discuss some implications of the obtained results. Since there
are no T-violating effects in two flavor neutrino case (in the limit
of vanshing $\theta_{13}$ or $\Delta m_{21}^2$, the three flavor
neutrino oscillations reduces to the two flavor ones), the T-violating
probability asymmetry can, in principle, provide a way to measure
$\theta_{13}$ and $\Delta m_{21}^2$.}
   

\preprint{\hepph{0108048}\\TUM-HEP-425/01}

\begin{document}

\section{Introduction}
\label{sec:intro}

T violation and CP violation in neutrino oscillations have lately been
extensively studied in the literature
\cite{akhm01,fish01,lately}. However, in most of  
these studies constant matter density was assumed in which only
fundamental (intrinsic) T violation is feasible. To learn more about
the effects of CP and T violation will be essential and necessary for
future experiments such as neutrino factories and other long baseline
neutrino oscillation experiments. Future experiments also offer to
study CPT violation. The measurement of CP, T, and CPT violation is
very important, not only because it will provide us with information
about neutrino properties, but also because it may have interesting
implications for physics at high energies.

This talk is based upon the work done by E.~Akhmedov, P.~Huber,
M.~Lindner, and T.~Ohlsson \cite{akhm01}.

\section{T-violating probability asymmetry}
\label{sec:Tviol}

We discuss the interplay of fundamental and matter-induced T violation
in three flavor neutrino oscillations in matter.\footnote{A T transformation is
a time reversal transformation.} T violation cannot be
directly experimentally tested, since this would mean changing the
direction of time. However, instead of studying neutrino oscillations
``backward'' in time, one can study them forward in time, but with the
initial and final neutrino flavors interchanged.
We define a measure of T violation in neutrino oscillations as the
following differences, which we will call the T-odd probability
differences:
\begin{equation}
\Delta P^T_{\alpha\beta} \equiv P(\nu_\alpha \to \nu_\beta) -
P(\nu_\beta \to \nu_\alpha),
\end{equation}
where $P(\nu_\alpha \to \nu_\beta)$ is the transition probability for
$\nu_\alpha \to \nu_\beta$. Furthermore, we will also denote $P(\nu_\alpha \to
\nu_\beta)$ by $P_{\alpha\beta}$.

For two neutrino flavors there are no T-viola\-ting effects simply
because $P_{e\mu} = P_{\mu e}$, which means that $\Delta P^T_{e\mu} =
0$. For three neutrino flavors the situation is more complicated and
we have to divide the problem into two separate cases: vacuum and matter.
In vacuum, we have CPT invariance, which means that we have T
violation if and only if we have CP violation. In matter, the
situation is different. Matter is both CP and CPT-asymmetric, since it
consists of particles (nucleons and electrons) and not of their
antiparticles or, in general, of unequal numbers of particles and
antiparticles. The matter density profiles are, of course, either symmetric or
asymmetric. Examples of the different types of matter density profiles
are shown in Figs.~\ref{fig:smdp} and \ref{fig:amdp}.
\FIGURE[!h]{
\begin{picture}(10,6)
\put(2,2){\vector(1,0){7}}
\put(9.2,1.85){{\small $L$}}
\put(2,2){\vector(0,1){4}}
\thicklines
\put(2,4){\line(1,0){6}}
\end{picture}
\begin{picture}(10,6)
\put(2,2){\vector(1,0){7}}
\put(9.2,1.85){{\small $L$}}
\put(2,2){\vector(0,1){4}}
\thicklines
\put(2,3){\line(2,1){3}}
\put(5,4.5){\line(2,-1){3}}
\end{picture}
\begin{picture}(10,6)
\put(2,2){\vector(1,0){7}}
\put(9.2,1.85){{\small $L$}}
\put(2,2){\vector(0,1){4}}
\thicklines
\put(5,2){\arc{6}{-3.14}{0}}
\end{picture}
\vspace{-1cm}
\caption{Examples of symmetric matter density profiles. Note that also
vacuum can be thought of as a symmetric matter density profile.}
\label{fig:smdp}
}

\FIGURE[!h]{
\begin{picture}(20,6)
\put(7,2){\vector(1,0){7}}
\put(14.2,1.85){{\small $L$}}
\put(7,2){\vector(0,1){4}}
\thicklines
\put(7,3){\line(1,0){3}}
\put(10,5){\line(1,0){3}}
\end{picture}
\vspace{-1cm}
\caption{Example of an asymmetric matter density profile (with two
layers of constant density).}
\label{fig:amdp}
}

The T-odd probability difference $\Delta P^T_{e\mu}$ (three neutrino
flavors) has been derived using perturbation theory to first order in
the parameter $\theta_{13}$ for arbitrary matter density profiles \cite{akhm01}
\begin{eqnarray}
\Delta P^T_{e\mu} &\simeq& - 2 s_{23} c_{23} \Im\left[ \beta^\ast
\left(A - C^\ast\right) \right] \nonumber\\
&\simeq& - 2 s_{13} s_{23} c_{23} \left(
\Delta - s_{12}^2 \delta \right) \nonumber\\
&\times& \Im\left[ e^{-i \delta_{CP}}
\beta^\ast \left(A_a - C_a\right) \right],
\label{eq:Tviol_main}
\end{eqnarray}
where $s_{ab} \equiv \sin \theta_{ab}$, $c_{ab} \equiv \cos
\theta_{ab}$, $\delta \equiv \frac{\Delta m_{21}^2}{2E_\nu}$, $\Delta
\equiv \frac{\Delta m_{31}^2}{2E_\nu}$, $A_a \equiv \alpha
\int_{t_0}^t \alpha^\ast f \, dt' + \beta \int_{t_0}^t \beta^\ast f \,
dt'$, and $C_a \equiv f \int_{t_0}^t \alpha f^\ast \, dt'$. Here
$\alpha = \alpha(t,t_0)$ and $\beta = \beta(t,t_0)$ are to be
determined from the solutions of the two flavor neutrino problem in
the $(1,2)$-sector (see Ref. \cite{akhm01} for details) and
\begin{eqnarray}
f &=& f(t,t_0) \nonumber\\
&\equiv& \exp \left\{ -i \int_{t_0}^t \left( \Delta -
\frac{1}{2} \left[ \delta + V(t') \right] \right) \, dt' \right\}, \nonumber
\end{eqnarray}
where $V(t) \simeq \frac{1}{\sqrt{2}} G_F \frac{1}{m_N} \rho(t)$ is
the matter potential with $G_F \simeq 1.16639 \cdot 10^{-23} \; {\rm
eV}^{-2}$ being the Fermi weak coupling constant, $m_N \simeq 939 \; {\rm
MeV}$ the nucleon mass, and $\rho(t)$ the matter density.
Note that formula~(\ref{eq:Tviol_main}) is valid only when $\theta_{13}$
and $\delta/\Delta$ are small parameters. In addition, it holds in
general that
$\Delta P^T_{e\mu} = \Delta P^T_{\mu\tau} = \Delta P^T_{\tau e}$
\cite{kras88}, {\it i.e.}, the T-odd probability differences are
cyclic in the indices and there is in fact only one independent T-odd
probability difference.
Furthermore, we have explicitly calculated the T-odd probability
difference $\Delta P^T_{e\mu}$ for
\begin{enumerate}
\item matter consisting of two layers (lengths $L_1$ and $L_2$) of
constant density (matter-induced potentials $V_1$ and $V_2$) and
\item an arbitrary matter density profile in the adiabatic
approximation.
\end{enumerate}

In the first case in the low energy regime ($\delta \gtrsim V_{1,2}$),
we obtain
\begin{eqnarray}
\Delta P_{e\mu}^T &\simeq& \cos \delta_{CP} \cdot 8 \underbrace{
s_{12} c_{12} s_{13} s_{23} c_{23} \frac{\sin(2\theta_1 - 2\theta_2)}{\sin
2\theta_{12}}}_{J_{\rm eff}} \nonumber\\
&\times& \left\{ \sin \omega_1 L_1 \sin \omega_2 L_2 \left[Y - \cos \left(
\Delta_1 L_1 + \Delta_2 L_2 \right) \right] \right\} \nonumber\\
&+& \sin \delta_{CP} \cdot 4 s_{13} s_{23} c_{23} \nonumber\\
&\times& X_1 \left[ Y - \cos
\left( \Delta_1 L_1 + \Delta_2 L_2 \right) \right],
\label{eq:Tviol_1}
\end{eqnarray}
where $\theta_1$ and $\theta_2$ are the matter mixing angles (in the
(1,2)-sector) in the first and second layers, respectively, $\Delta_a
= \Delta - \frac{1}{2} \left( \delta + V_a \right)$ \quad ($a = 1,2$),
\begin{eqnarray}
Y &\equiv& \cos \omega_1 L_1 \cos \omega_2 L_2 \nonumber\\
&-& \sin \omega_1 L_1 \sin \omega_2 L_2 \cos(2\theta_1 - 2\theta_2),
\nonumber\\
X_1 &\equiv& \sin \omega_1 L_1 \cos \omega_2 L_2 \sin 2\theta_1 \nonumber\\
&+& \sin \omega_2 L_2 \cos \omega_1 L_1 \sin 2\theta_2 \nonumber
\end{eqnarray}
with
$$
\omega_a \equiv \frac{1}{2} \sqrt{\left(\cos 2\theta_{12} \delta -
V_a\right)^2 + \sin^2 2\theta_{12} \delta^2}
$$
($a = 1,2$) and $J_{\rm eff}$ is an effective Jarlskog invariant similar to the
usual Jarlskog invariant \cite{jarl85}
$$
J \equiv s_{12} c_{12} s_{13} c_{13}^2 s_{23} c_{23} \sin \delta_{CP}.
$$
In formula~(\ref{eq:Tviol_1}), the $\cos \delta_{CP}$ and $\sin
\delta_{CP}$ terms describe matter-induced (extrinsic) and fundamental
(intrinsic) T violation, respectively.

In the second case in the regime in which the oscillations governed by
large $\Delta$ are fast and therefore can be averaged over ($V(t)
\lesssim \delta \ll \Delta$), we obtain
\begin{eqnarray}
\Delta P^T_{e\mu} &\simeq& \cos \delta_{CP} \cdot 4 \underbrace{s_{12} c_{12}
s_{13} s_{23} c_{23} \frac{\sin(2\theta - 2\theta_0)}{\sin
2\theta_{12}}}_{J_{\rm eff}} \nonumber\\
&\times& \cos^2 \Phi \nonumber\\
&+& \sin \delta_{CP} \cdot 2 s_{13} s_{23} c_{23} \sin(\theta +
\theta_0) \cos(\theta - \theta_0) \nonumber\\
&\times& \sin 2\Phi,
\end{eqnarray}
where $\theta_0$ and $\theta$ are the matter mixing angles (in the
(1,2)-sector) at the initial and final points of the neutrino
evolution, $t_0$ and $t$, respectively, $\Phi \equiv \int_{t_0}^t
\omega(t') \, dt'$ with
$$
\omega(t) \equiv \frac{1}{2} \sqrt{\left(\cos 2\theta_{12} \delta -
V(t)\right)^2 + \sin^2 2\theta_{12} \delta^2}.
$$
Note that the $\cos \delta_{CP}$ term can again be written in terms of
the effective Jarlskog invariant.

\section{Precision of the analytical approximation}
\label{sec:precision}

Next we discuss the precision of the approximative analytical
formula for the first case, {\it i.e.}, matter consisting of two
layers of constant density. In order to do so, we have to
distinguish between two cases: $L/E \lesssim 10^4 \; {\rm km/GeV}$
and $L/E \gtrsim 10^4 \; {\rm km/GeV}$.

The first case is shown in Fig.~\ref{fig:compphase}, whereas the
second case is shown in Fig.~\ref{fig:compaver}. In the first case,
the oscillating structure of the T-odd probability difference $\Delta
P^T_{e\mu}$ can be resolved. The amplitude of $\Delta P^T_{e\mu}$ is
reproduced very well; however, there is an error in the phase that
is accumulating with the baseline length $L$ as well as with growing
$\theta_{13}$ and $\Delta m_{21}^2$.
\FIGURE[!h]{
\includegraphics[width=0.25 \textwidth,angle=-90]{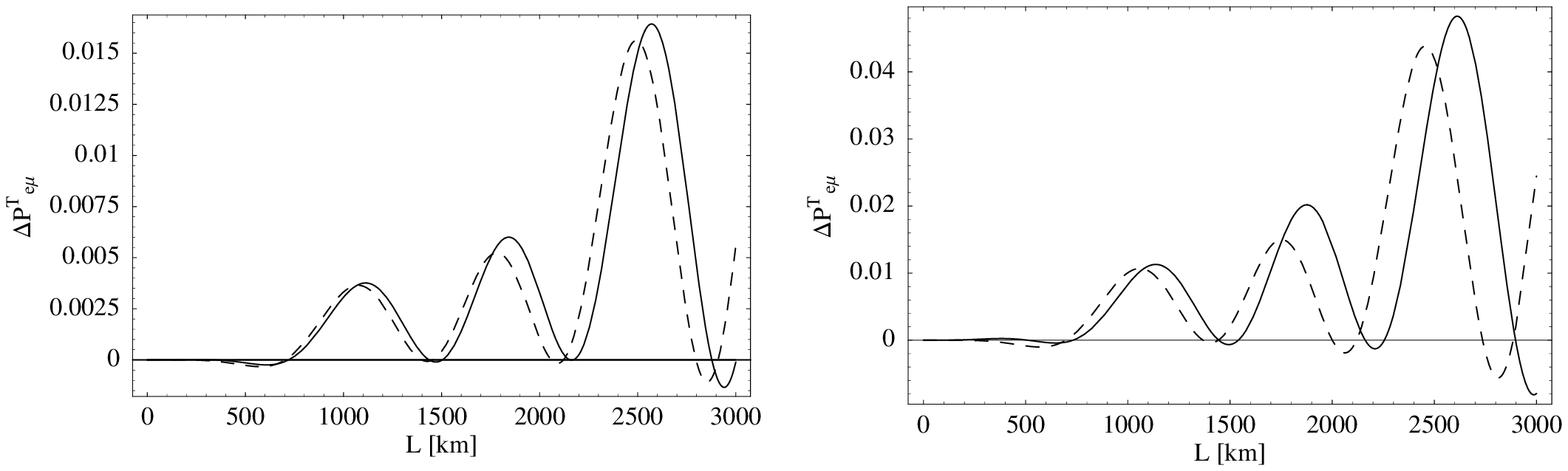}
\caption{The T-odd probability difference $\Delta P^T_{e\mu}$ for two
layers of widths $L_1 = L_2 = L/2$ and densities $1 \; {\rm g/cm^3}$
and $3 \; {\rm g/cm^3}$ as a function of the baseline length $L$. Solid
curves show analytical results, whereas dashed curves show numerical
results. Left plot: $\theta_{13} = 0.1$ and $\Delta m_{21}^2 = 5 \cdot
10^{-5} \; {\rm eV}^2$. Right plot: $\theta_{13} = 0.16$ and $\Delta
m_{21}^2 = 2 \cdot 10^{-4} \; {\rm eV}^2$. Remaining parameters:
$\Delta m_{31}^2 = 3.5 \cdot 10^{-3} \; {\rm eV}^2$, $\theta_{12} =
0.56$, $\theta_{23} = \pi/4$, $\delta_{CP} = 0$, and $E = 1 \; {\rm
GeV}$. The figure has been taken from Ref. \cite{akhm01}.}
\label{fig:compphase}
}
In the second case, the oscillations governed by the $\Delta m_{31}^2$
are very fast.
\FIGURE[!h]{
\includegraphics[width=0.25 \textwidth,angle=-90]{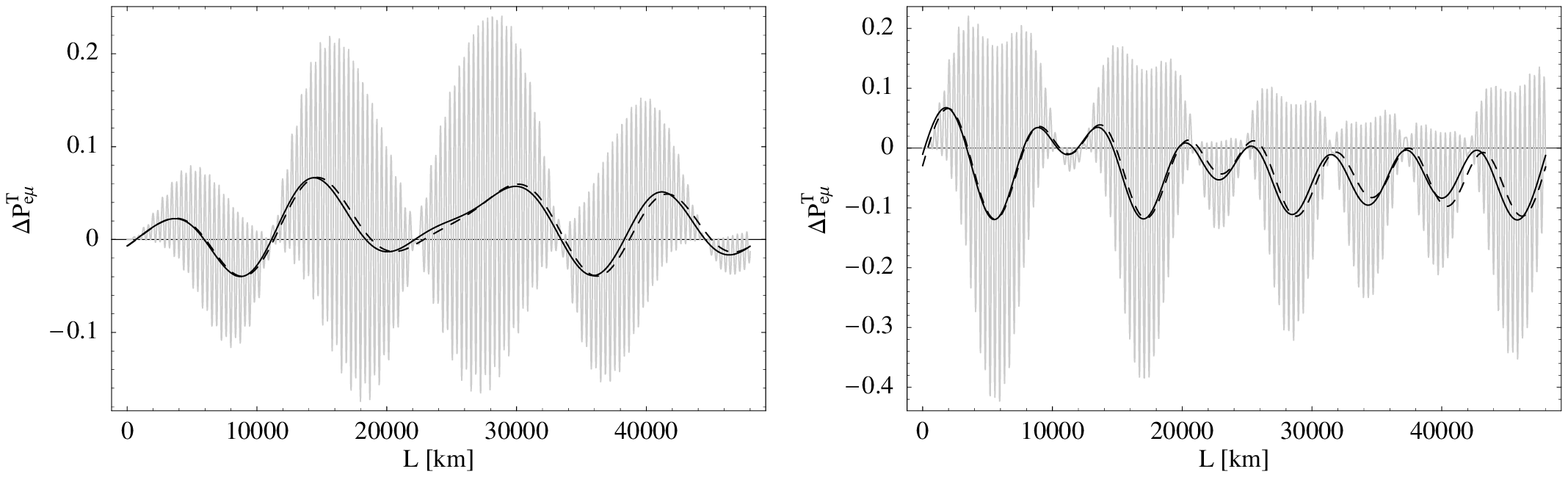}
\caption{The same as in Fig.~\ref{fig:compaver} except that the
densities are $0$ and $6.4 \; {\rm g/cm^3}$ and $E = 0.5 \; {\rm
GeV}$. The grey curves are analytical results, whereas the black solid
and dashed curves show the results averaged over fast oscillations of
the analytical and numerical calculations, respectively. The figure
has been taken from Ref. \cite{akhm01}.}
\label{fig:compaver}
}
The left plot uses the same parameter values as in Fig. 3a of
P.M. Fishbane and P. Kaus \cite{fish01}, whereas the right plot uses
larger values of $\theta_{13}$ and $\Delta m_{21}^2$.

What about T violation in future terrestrial neutrino oscillation
experiments including neutrino factory experiments? Will it give rise
to sizeable effects that are measurable?
We studied several different experimental setups for a neutrino
factory with a beam energy of 50 GeV. Furthermore, we assumed $2 \cdot
10^{21}$ muon decays and a detector mass of 40 kton. (See
Ref. \cite{akhm01} for further details.) We investigated two different
two-layer matter density profiles with densities $1 \; {\rm g/cm}^3$,
$3 \; {\rm g/cm}^3$ and $3 \; {\rm g/cm}^3$, $3.3 \; {\rm g/cm}^3$,
respectively. The first one could correspond to a sea-earth scenario and
the second one to a very long baseline experiment in which there should
be density perturbations. We simulated these scenarios and performed
fits to the obtained event rates. In addition, we compared these
simulations for symmetrized versions of the corresponding matter
density profiles. The symmetrized matter density profiles are modeled
by replacing the transition probabilities with the symmetrized ones:
\begin{equation}
P_S = \frac{1}{2} \left( P_{\rm dir} + P_{\rm rev}\right),
\end{equation}
where $P_{\rm dir}$ and $P_{\rm rev}$ are the transition probabilities
originating from the neutrino propagation in the (direct)
matter density profile and the ``reverse'' matter density profile,
respectively.\footnote{Time reversal implies that the matter density
profile has to be traversed in the opposite direction.}
Thus, the simulations are only sensitive to errors induced by the
asymmetry of the matter density profile. The difference of the minimal
values of the $\chi^2$ functions for the asymmetric and symmetrized
matter density profiles is a measure of matter-induced T violation. 

Our simulations show that the T-violating effects can be ``quite
sizeable'' for the sea-earth matter density profile; however, only
for $L \gtrsim 1000 \; {\rm km}$, which cannot be realized on Earth.
The qualitative statements of the simulations do not change very much
if one changes the value of $\delta_{CP}$ in the fits. For the matter
density profile with 10\% density perturbations, the matter-induced
T-violating effects are small for any baseline.

\section{Summary and conclusions}
\label{sec:SC}

In summary, approximative analytical formulas for the T-odd
probability differences $\Delta P^T_{\alpha\beta}$ for an arbitrary
matter density profile have been derived using perturbation theory.

Our main conclusions are the following: 
\begin{itemize}
\item T-violating effects can be considered as a measure of genuine
three-flavorness.
\item For terrestial experiments matter-induced T-violating effects
can safely be ignored.
\item Asymmetric matter effects cannot hinder the determination of the
fundamental CP and T-violating phase $\delta_{CP}$ in long baseline
experiments.
\end{itemize}

\acknowledgments

I would like to thank my co-workers Evgeny Akhmedov, Patrick Huber,
and Manfred Lindner for fruitful collaboration and Walter Winter for
proof-reading this proceeding.

This work was supported by the Swedish Foundation for International
Cooperation in Research and Higher Education (STINT), the Wenner-Gren
Foundations, and the ``Sonderforschungsbereich 375 f{\"u}r
Astro-Teilchenphysik der Deutschen Forsch\-ungsgemeinschaft''.

\end{document}